\documentclass[preprint,nofootinbib,amsmath,amssymb]{revtex4}

\usepackage[dvips]{graphicx}

\begin{document}

\title{Constraints on the Phase Plane of the Dark Energy Equation of State}

\author{Chien-Wen Chen}
\email{f90222025@ntu.edu.tw} %
\affiliation{Department of Physics, National Taiwan University, Taipei 10617, Taiwan, R.O.C.} %
\affiliation{Leung Center for Cosmology and Particle Astrophysics
(LeCosPA), National Taiwan University, Taipei 10617, Taiwan, R.O.C.}

\author{Pisin Chen}
\affiliation{Kavli Institute for Particle Astrophysics and
Cosmology, SLAC National Accelerator Laboratory, Menlo Park, CA
94025, U.S.A.} \affiliation{LeCosPA, Department of Physics, and
Graduate Institute of Astrophysics,
National Taiwan University, Taipei 10617, Taiwan, R.O.C.} %

\author{Je-An Gu}
\affiliation{LeCosPA, National Taiwan University, Taipei 10617,
Taiwan, R.O.C. }

\begin{abstract}
Classification of dark energy models in the plane of $w$ and
$w^\prime$, where $w$ is the dark energy equation of state and
$w^\prime$ its time-derivative in units of the Hubble time, has been
studied in the literature. We take the current SN Ia, CMB and BAO
data, invoke a widely used parametrization of the dark energy
equation of state, and obtain the constraints on the $w \textrm{--}
w^\prime$ plane. We find that dark energy models including the
cosmological constant, phantom, non-phantom barotropic fluids, and
monotonic up-rolling quintessence are ruled out at the $68.3\%$
confidence level based on the current observational data.
Down-rolling quintessence, including the thawing and the freezing
models, is consistent with the current observations. All the
above-mentioned models are still consistent with the data at the
$95.4\%$ confidence level.
\end{abstract}

\pacs{95.36.+x}

\maketitle

\section{Introduction\label{sec:introduction}}
Compelling evidence from different types of observation shows that
the expansion of the universe is accelerating at late times
(see~\cite{Frieman:2008sn} for a review). Within the framework of
general relativity, this indicates that there should exist an energy
source with significant negative pressure, termed dark energy, to
drive this acceleration. The nature of dark energy is generally
regarded as one of the most tantalizing problems in cosmology. Many
dark energy models have been proposed and studied
(see~\cite{Frieman:2008sn,Chen:2009xv}, and references therein).
While the cosmological constant remains the simplest realization of
dark energy, current observations do not rule out the possibility of
time-evolving dark
energy~\cite{Frieman:2008sn}$\textrm{--}$\cite{Albrecht:2006um}.

In the pursuit of revealing the nature of dark energy, cosmological
observations serve to constrain the behavior of dark energy.
Theoretical studies, on the other hand, should determine whether
dark energy models can be distinguished by their observational
consequences. The ratio of pressure to energy density for dark
energy, the equation of state $w=p/\rho$, is the characteristic of
how the energy density evolves with time. The cosmological constant
relates to the constant equation of state $w=-1$, while other dark
energy models generally have time-evolving $w$. The time-derivative
of $w$ in units of the Hubble time, $w'=dw/dlna$, characterizes the
dynamical behavior of the equation of state. Studies of the
dynamical behaviors and classification of dark energy models in the
$w\textrm{--}w'$ phase plane have been carried
out~\cite{Caldwell:2005tm}$\textrm{--}$\cite{Linder:2008ya}. It is
found that different dark energy models are bounded in different
sectors in the $w\textrm{--}w'$ plane.

In this paper, on the one hand, we gather the bounds for various
dark energy models in the $w \textrm{--} w^\prime$ plane. On the
other hand, we obtain the constraints on the $w \textrm{--}
w^\prime$ plane in the redshift region $0<z<1$, by adopting a widely
used
parametrization~\cite{Chevallier:2000qy,Linder:2002et,Albrecht:2006um},
$w(z) = w_0 + w_a(1-a) = w_0 + w_a z/(1+z)$, based on the current
observational data. The data set we use includes the recently
compiled ``Constitution set'' of Type Ia supernovae (SN Ia)
data~\cite{Riess:2004nr}$\textrm{--}$\cite{Hicken:2009dk}, the
cosmic microwave background (CMB) measurement from the five-year
Wilkinson Microwave Anisotropy Probe (WMAP)~\cite{Komatsu:2008hk},
and the baryon acoustic oscillation (BAO) measurement from the Sloan
Digital Sky Survey (SDSS)~\cite{Eisenstein:2005su} and the 2dF
Galaxy Redshift Survey ($2$dFGRS)~\cite{Percival:2007yw}. We then
compare the dark energy models with the constraints on the $w
\textrm{--} w^\prime$ plane for $0<z<1$. The work close to ours is
that of Barger et al.~\cite{Barger:2005sb}, in which they used the
earlier data set and examined the dark energy models in the $w_0
\textrm{--} w_a$ plane only at the redshift $z=1$ .

\section{Classification of dark energy models}
{\it Quintessence}

The quintessence
model~\cite{Caldwell:1998ii}$\textrm{--}$\cite{Boyle:2001du}, which
invokes a time-varying scalar field, generally allows its energy
density and equation of state to evolve with time, and has $w>-1$.
The equation of motion of the quintessence field is
$\ddot\phi+3H\dot\phi+V_{,\phi}=0$, where $H=\dot a/a$ is the Hubble
expansion rate, and $V_{,\phi}=dV/d\phi$. In terms of $w$ and
$w^\prime$, the equation of motion can be written
as~\cite{Steinhardt:1999nw}
\begin{equation}\label{eq:q1}
\mp \frac{V_{,\phi}}{V} =
\sqrt{\frac{3(1+w)}{\Omega_\phi(a)}}\left[1 + \frac{1}{6}
\frac{d\ln(x_q)}{d\ln(a)}\right],
\end{equation}
where the minus sign corresponds to $\dot\phi>0$ and the plus sign
to the opposite, $\Omega_\phi(a)$ is the dimensionless energy
density of the quintessence field, and $x_q = (1+w)/(1-w)$. For the
down-rolling quintessence field ($\dot V<0$), the left-hand side of
Eq.~(\ref{eq:q1}) is positive, and the bound of $w$ and $w^\prime$
can be obtained as $w^\prime >
-3(1-w)(1+w)$~\cite{Scherrer:2005je,Chiba:2005tj}. The up-rolling
quintessence field ($\dot V>0$) takes the other side,
$w^\prime<-3(1-w)(1+w)$. The bound of the tracker
quintessence~\cite{Zlatev:1998tr} is obtained
in~\cite{Scherrer:2005je,Chiba:2005tj}. However, strong acceleration
today, with $w\lesssim-0.7$, requires the breakdown of
tracking~\cite{Linder:2006sv}. The bound should only apply to the
high redshift~\cite{Linder:2006sv}, $z\gg1$, which is not the region
of interest in this paper. A conjectured limit of quintessence has
been proposed in~\cite{Linder:2006sv} as $V/(-V_{,\phi})<M_P$, where
$M_P$ is the Plank mass. However, the physical origin of this limit
is not clear~\cite{Linder:2006sv}. We therefore do not impose this
constraint on the quintessence model. Caldwell and Linder identified
two categories of quintessence models, ``thawing'' and ``freezing'',
based on their dynamical behavior~\cite{Caldwell:2005tm}. For the
thawing models, the equation of state is $w\approx -1$ at early
times, but grows less negative with time as $w'>0$. The bounds of
the thawing models are $(1+w)<w'<3(1+w)$. For the freezing models,
initially the equation of state is $w>-1$ with $w'<0$, but the field
is frozen at late times where $w \to -1$ and $w'\to 0$. The bounds
of the freezing models are $3w(1+w)<w'<0.2w(1+w)$. Note that the
upper bound for the freezing models is only valid for $z<1$.

{\it Phantom}

The phantom model has negative kinetic energy and the equation of
state $w<-1$~\cite{Caldwell:1999ew}. The equation of motion of the
phantom field is $\ddot\phi+3H\dot\phi-V_{,\phi}=0$. In terms of $w$
and $w^\prime$, the equation of motion can be written
as~\cite{Kujat:2006vj}
\begin{equation}\label{eq:p1}
\pm \frac{V_{,\phi}}{V} =
\sqrt{\frac{-3(1+w)}{\Omega_\phi(a)}}\left[1 + \frac{1}{6}
\frac{d\ln(x_p)}{d\ln(a)}\right],
\end{equation}
where the plus sign corresponds to $\dot\phi>0$ and the minus sign
to the opposite, $\Omega_\phi(a)$ is the dimensionless energy
density of the phantom field, and $x_p = -(1+w)/(1-w)$. For the
up-rolling phantom field ($\dot V>0$), the left-hand side of
Eq.~(\ref{eq:p1}) is positive, and the bound of $w$ and $w^\prime$
can be obtained as $w^\prime < -3(1-w)(1+w)$. The down-rolling
phantom field ($\dot V<0$) takes the other side
$w^\prime>-3(1-w)(1+w)$. Note that Eq.~(\ref{eq:p1}) and the bounds
are different from those obtained in~\cite{Chiba:2005tj}.

{\it Barotropic fluids}

Barotropic fluids are those for which the pressure is an explicit
function of the energy density, $p=f(\rho)$
(see~\cite{Linder:2008ya} and references therein). The expression
for $w'$ can be written as~\cite{Scherrer:2005je,Linder:2008ya}
\begin{equation}
\label{baro} w^\prime = -3(1+w)\left(\frac{dp}{d\rho} - w\right).
\end{equation}
The sound speed for a barotropic fluid is given by $c_s^2=dp/d\rho$.
To ensure stability, we must have $c_s^2 \geq 0$, which gives the
bound $w' \leq 3w(1+w)$ for non-phantom ($w>-1$) barotropic
fluids~\cite{Scherrer:2005je,Linder:2008ya}. For causality, we
further require $c_s^2 \leq 1$~\cite{Ellis:2007ic}, which gives the
bound $w' \geq -3(1+w)(1-w)$ for $w>-1$~\cite{Linder:2008ya}.

\begin{figure}[ph]
\begin{center}
        \mbox{\includegraphics[width=0.65\linewidth]{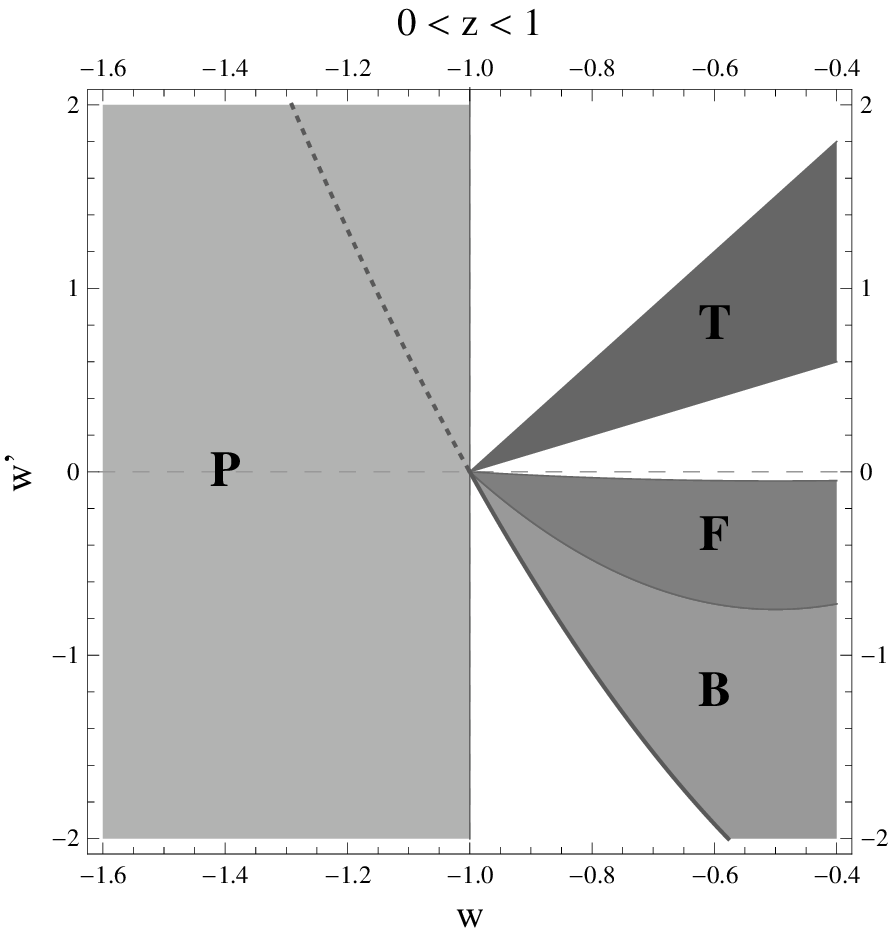}}
\vspace*{8pt} \caption{Classification of dark energy models in the
$w \textrm{--} w^\prime$ plane. Models are separated by the solid
curves. The symbols ``T'', ``F'', ``B'', and ``P'' denote the
``thawing'', ``freezing'', ``non-phantom barotropic'', and
``phantom'' models, respectively. The quintessence models correspond
to the region for $w>-1$ . The cosmological constant corresponds to
the point $(-1,0)$. The bold solid curve is both the lower bound for
the non-phantom barotropic models and the bound that separates the
down-rolling and up-rolling quintessence models (down-rolling takes
the upper side). The dotted curve is the bound that separates the
down-rolling and up-rolling phantom models (up-rolling takes the
lower side).} \label{fig1}
\end{center}
\end{figure}

The classification of the above-mentioned dark energy models in the
$w \textrm{--} w^\prime$ plane is shown in Fig.~\ref{fig1}. Note
that all of the bounds are valid at late times for $0<z<1$.

\section{Constraints on the $w \textrm{--} w^\prime$ plane}

\subsection{Observational data}
We use the combined data set from three types of observations
including the SN Ia observation, the CMB measurement, and the BAO
measurement. We assume that the universe is flat in this paper.

We use the Constitution set of SN Ia data compiled by Hicken et
al.~\cite{Riess:2004nr}$\textrm{--}$\cite{Hicken:2009dk}, which
provides the information of the luminosity distance and the
redshift. The SN Ia samples lie in the redshift region $0<z<1.55$.
The luminosity distance-redshift relation is given by
\begin{equation} \label{eq:DL}
d_L(z)=(1+z)\int_0^{z}\frac{dz'}{H(z')}.
\end{equation}
We use the CMB shift parameter measured by the five-year WMAP
observation \cite{Komatsu:2008hk},
\begin{equation} \label{eq:R}
 R = \sqrt{\Omega _m H_0^2} \int_0^{1090.04}\frac{dz}{H(z)}=1.710\pm 0.019,
 \end{equation}
where $H_0$ is the Hubble constant and $\Omega_m$ is the
dimensionless matter density at present. We use the BAO measurement
from the joint analysis of the SDSS and 2dFGRS
data~\cite{Eisenstein:2005su,Percival:2007yw}, which gives
$D_V(0.35)/D_V(0.2) = 1.812\pm 0.060$, where
\begin{equation}\label{eq:BAO1}
D_V(z_\textsc{bao}) = \left[(1+z_\textsc{bao})^2
D_A^2(z_\textsc{bao})
\frac{z_\textsc{bao}}{H(z_\textsc{bao})}\right]^{1/3},
\end{equation}
and $D_A(z)$ is the angular diameter distance,
\begin{equation} \label{eq:DA}
D_A(z)=\frac{1}{1+z}\int_0^{z}\frac{dz'}{H(z')}.
\end{equation}

\begin{figure}[ph]
\begin{center}
        \mbox{\includegraphics[width=0.65\linewidth]{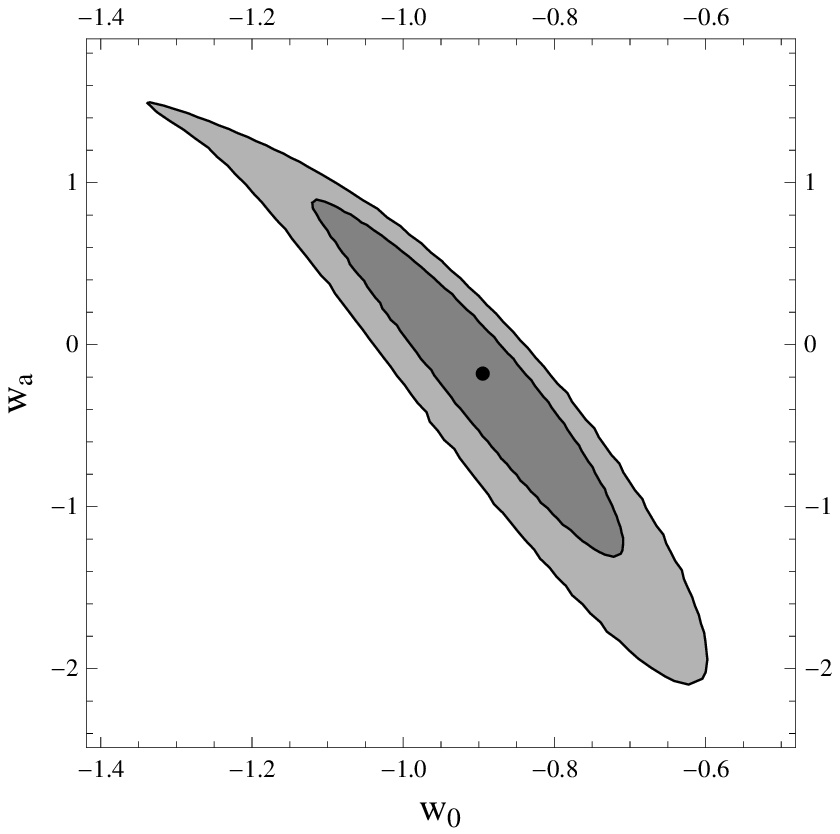}}
\vspace*{8pt} \caption{The two-dimensional constraint of
$w_0\textrm{--}w_a$ based on the combined data set including the
Constitution set of SN Ia data, the CMB measurement from the
five-year WMAP, and the BAO measurement from the SDSS and 2dFGRS.
The dark and the light gray areas correspond to the $68.3\%$ and the
$95.4\%$ confidence regions, respectively. The black point denotes
the best-fit values.} \label{fig2}
\end{center}
\end{figure}

To obtain the constraints on the $w \textrm{--} w^\prime$ plane, we
invoke a broadly used form of parametrization of the equation of
state~\cite{Chevallier:2000qy,Linder:2002et,Albrecht:2006um},
\begin{equation}
 w(z) = w_0 + w_a
(1-a) = w_0 + w_a z/(1+z) \,.  \label{eq:w0wa-parametrization}
\end{equation}The constraint of $w_0$, $w_a$ and $\Omega_m$ is obtained by fitting the three parameters to this combined data set. The estimate of the parameters are found to be
$  w_0=-0.89^{+0.12}_{-0.14}, \quad w_a=-0.18^{+0.71}_{-0.74}, \quad
\Omega_m= 0.25^{+0.03}_{-0.02}. $ The two-dimensional constraint of
$w_0 \textrm{--} w_a$ is obtained and shown in Fig.~\ref{fig2}.

\subsection{Results of the constraints on the $w \textrm{--} w^\prime$
plane}\label{phase_constraints} We reconstruct the the $w\textrm{--}
w^\prime$ plane via Eq.~(\ref{eq:w0wa-parametrization}) and
\begin{equation} w'(z)=-a w_a=-w_a/(1+z),  \label{eq:w_prime}
\end{equation}
at late times for $0<z<1$. At each redshift, the two-dimensional
constraint is obtained by converting the points on the boundaries of
the confidence regions in the $w_0 \textrm{--} w_a$ plane to the
corresponding points in the $w \textrm{--} w^\prime$ plane,
following Eq.~(\ref{eq:w0wa-parametrization}) and
Eq.~(\ref{eq:w_prime}), for the $68.3\%$ and the $95.4\%$ confidence
regions respectively.\footnote{For the one-dimensional error
propagation, following the reconstruction equations, the variance
propagation is $Var(w)= \langle(w-\langle w\rangle)^2\rangle =
\langle[w_0-\langle w_0\rangle+(1-a)(w_a-\langle
w_a\rangle)]^2\rangle=Var(w_0)+(1-a)^2Var(w_a)+2(1-a)Cov(w_a,w_a)$
and $Var(w^\prime)=(-a)^2Var(w_a)$. } Since the transformation
between ($w, w^\prime$) and ($w_0,w_a$) is linear, each point inside
a confidence region in the $w_0 \textrm{--} w_a$ plane gives a
distinct point inside the corresponding confidence region in the the
$w \textrm{--} w^\prime$ plane.

\begin{figure}[ph]
\begin{center}
        \mbox{\includegraphics[width=1\linewidth]{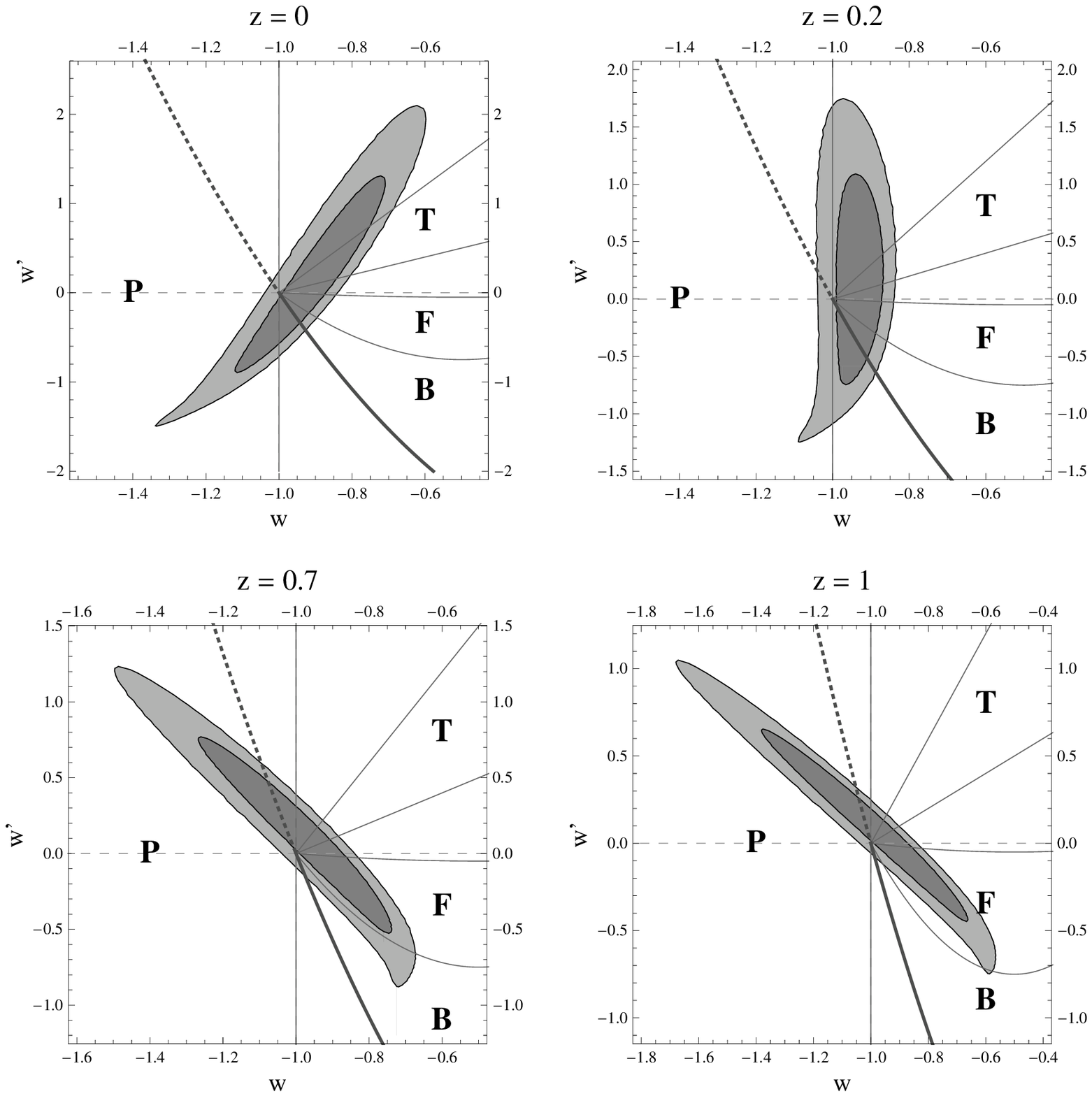}}
\vspace*{8pt} \caption{Samples of the constraints on the $w
\textrm{--} w^\prime$ plane at redshifts $z=0$, $0.2$, $0.7$ and
$1$. The dark and the light gray areas correspond to the $68.3\%$
and the $95.4\%$ confidence regions, respectively. See the caption
in Fig.~\ref{fig1} for the description of the regions to that the
models belong. The cosmological constant lie outside the $68.3\%$
confidence region at all of the redshifts. The down-rolling phantom
models lie outside the $68.3\%$ confidence region at $z=0$ and
$z=0.2$. All the phantom models lie outside the $68.3\%$ confidence
region at $z=0.2$. Both the up-rolling quintessence models and the
non-phantom barotropic fluids lie outside the $68.3\%$ confidence
region at $z=0.7$ and $z=1$. The down-rolling quintessence models
including the thawing and the freezing models overlap with the
$68.3\%$ confidence region at the four redshifts. All of the models
overlap with the $95.4\%$ confidence region at the four redshifts.}
\label{fig3}
\end{center}
\end{figure}

In the $w \textrm{--} w^\prime$ plane, we find that the cosmological
constant, the phantom models, the up-rolling quintessence models,
and the non-phantom barotropic fluids lie outside the $68.3\%$
confidence region in the redshift regions $0<z<1$, $0.18<z<0.22$,
$0.4<z<1$ and $0.7<z<1$, respectively. This shows that the four
models are ruled out at the $68.3\%$ confidence level. On the
contrary, the down-rolling quintessence models, including the
thawing and the freezing models, overlap with the $68.3\%$
confidence region for $0<z<1$. All of the models overlap with the
$95.4\%$ confidence region for $0<z<1$. Samples of the constraints
on the $w \textrm{--} w^\prime$ plane at redshifts $z=0$, $0.2$,
$0.7$ and $1$, together with the models, are shown in
Fig.~\ref{fig3}.

\section{Test of the method}\label{test}
In Sec.~\ref{phase_constraints}, we invoked a criterion that a model
is ruled out at the $68.3\%$ confidence level if for some redshift
the model's corresponding sector in the $w \textrm{--} w^\prime$
plane does not overlap with the confidence region at all. To test
the validity of this criterion and to address the concern about the
inherent bias of the parametrization against certain models, we
perform a Monte Carlo test of our method. The criterion is invalid
if for some redshift the resulting $68.3\%$ region from the Monte
Carlo realization of the fiducial model does not overlap at all with
the model's corresponding sector in the $w \textrm{--} w^\prime$
plane. We pick one or two fiducial models for each model category to
test our method.

\begin{figure}[ph]
\begin{center}
        \mbox{\includegraphics[width=0.7\linewidth]{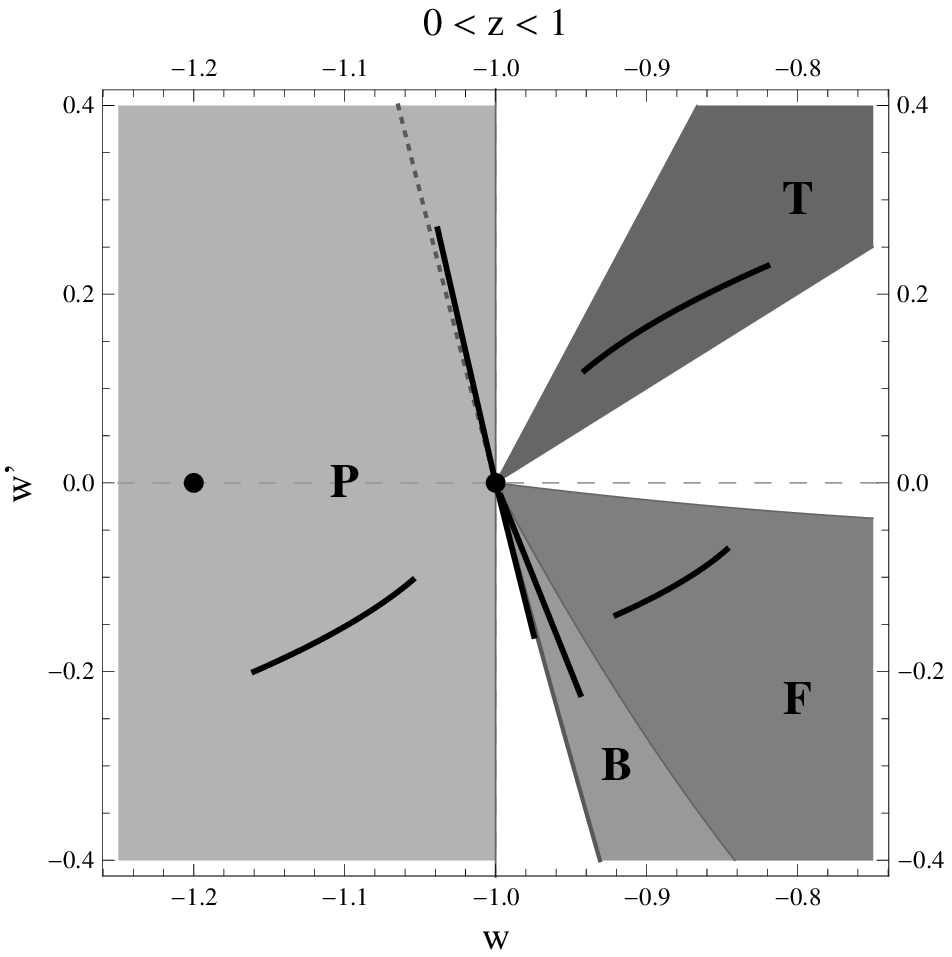}}
\vspace*{8pt} \caption{Trajectories of the fiducial models for the
test in the redshift region $0<z<1$. See the text in Sec.~\ref{test}
for the description of the models. The trajectories are shown by the
black points and the black curves.} \label{fig4}
\end{center}
\end{figure}

The fiducial models used in the test include {\it cosmological
constant}: $w(z)=-1$, {\it thawing}:
$w(z)=-0.82+0.23~lna+0.08~(lna)^2$, {\it freezing}:
$w(z)=-0.92-0.14~lna-0.05~(lna)^2$, {\it up-rolling quintessence}:
$w(z)=-1+0.0003~a^{-6.4}$, {\it up-rolling phantom}: (a)
$w(z)=-1.2$, (b) $w(z)=-1.16-0.2~lna-0.07~(lna)^2$, {\it
down-rolling phantom}: $w(z)=-1-0.0003~a^{-7}$, and {\it
none-phantom barotropic fluids}: $w(z)=-1+0.0035~a^{-4}$. For all
models, $\Omega_m$ is 0.25 and $w(z>1.55)$ is equal to $w(z=1.55)$.
All models have $w(z)<-0.8$ for strong acceleration today. The
trajectories of the models in the $w \textrm{--} w^\prime$ plane are
shown in Fig.~\ref{fig4} in the redshift region $0<z<1$.

In the test, we realize each fiducial model by simulating the SN Ia,
the BAO and the CMB data assuming current data quality. 1000 sets of
simulated data are generated and fitted with the three parameters
$w_0$, $w_a$ and $\Omega_m$. The values of $w_0$ and $w_a$ are
converted to $w$ and $w^\prime$ via
Eq.~(\ref{eq:w0wa-parametrization}) and Eq.~(\ref{eq:w_prime}) at
each redshift. The 683 and the 954 of the 1000 Monte Carlo
realizations, representing the $68.3\%$ and the $95.4\%$ region, are
selected via their chi-square from the fiducial model.

\begin{figure*}[ph]
\begin{center}
        \mbox{\includegraphics[width=0.71\linewidth]{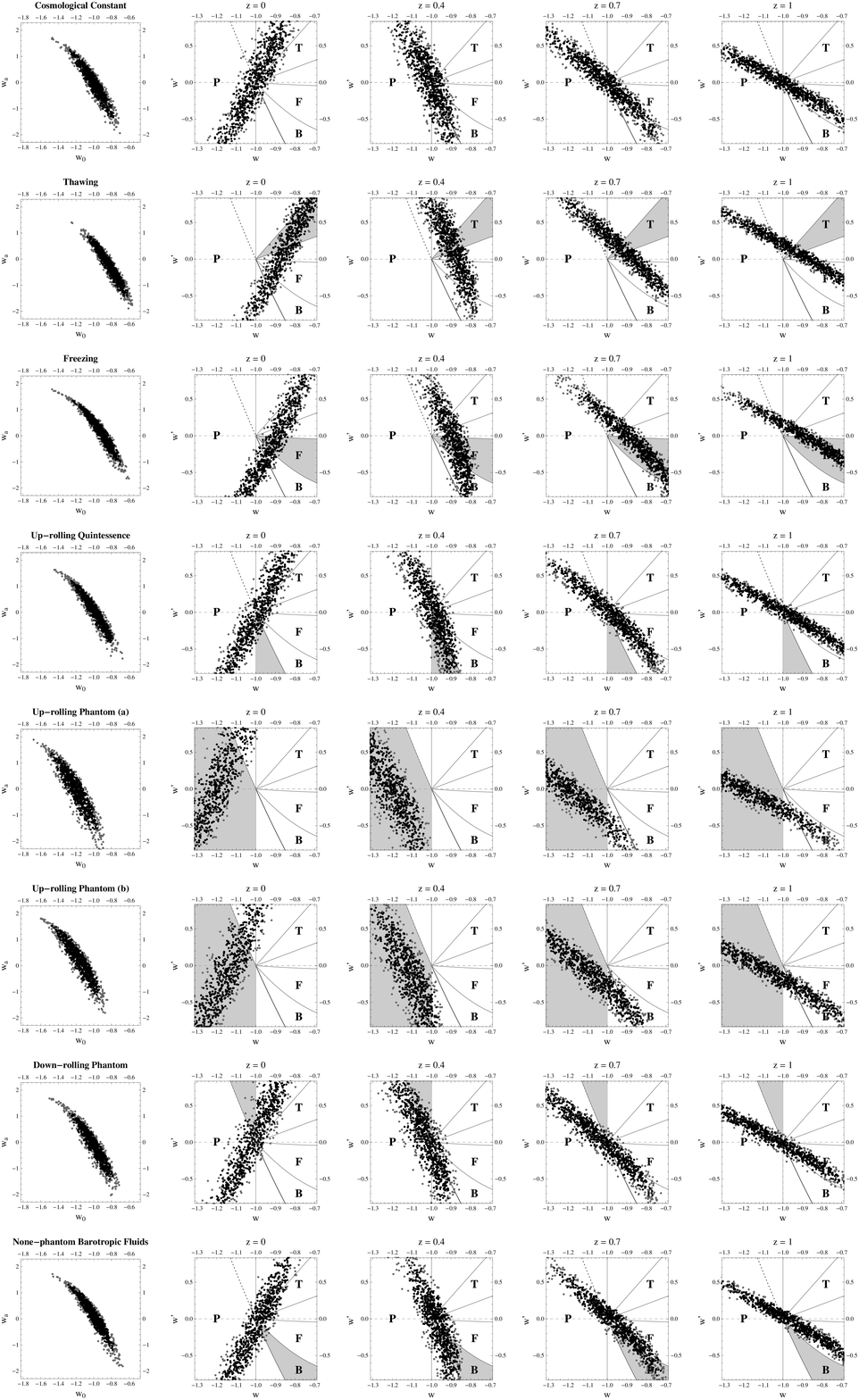}}
\vspace*{8pt} \caption{Samples of the Monte Carlo test results of
the method. From the first to the eighth row are the results
corresponding to the fiducial models of cosmological constant,
thawing, freezing, up-rolling quintessence, up-rolling phantom (a)
and (b), down-rolling phantom and non-phantom barotropic fluids,
respectively. The first column is the $w_0 \textrm{--} w_a$ plane
and the rest four are the $w \textrm{--} w^\prime$ plane at
redshifts $z=0$, $0.4$ $0.7$ and $1$. The $68.3\%$ Monte Carlo
realized region is represented by the black points, while the
$95.4\%$ by the black and the dark grey points. The corresponding
sectors of the fiducial models are filled with the light grey color.
See Sec.~\ref{test} for discussion.} \label{fig5}
\end{center}
\end{figure*}

As a result, the criterion passes the test for all the models, that
is, for each model the corresponding sector overlap with the
$68.3\%$ Monte Carlo realized region in the $w \textrm{--} w^\prime$
plane in the redshift region $0<z<1$. Samples of the test results at
redshifts $z=0$, $0.4$, $0.7$ and $1$ are shown in Fig.~\ref{fig5}.
It is seen that applying the criterion up to $z=1$ might be pushing
it to the limit, especially for the up-rolling quintessence,
down-rolling phantom and the non-phantom barotropic fluid cases. Yet
the conclusion that the three models are ruled out at the $68.3\%$
confidence level in Sec.~\ref{phase_constraints} still holds if we
apply the the criterion only for $0<z<0.7$.

\section{Conclusion and discussion}
Applying the bounds for various dark energy models in the $w
\textrm{--} w^\prime$ plane for redshift $0<z<1$, we find that
several models including the cosmological constant, phantom,
non-phantom barotropic fluids, and monotonic up-rolling quintessence
are ruled out at the $68.3\%$ confidence level based on the current
observational data. On the other hand, down-rolling quintessence,
including the thawing and the freezing models, is consistent with
the current observations. All the models are still consistent with
the data at the $95.4\%$ confidence level. Using the same SN Ia data
set, Shafieloo et al.~\cite{Shafieloo:2009ti} and Huang et
al.~\cite{Huang:2009rf} also found the cosmological constant
inconsistent with the data at the $68.3\%$ confidence level. Barger
et al.~\cite{Barger:2005sb} found the non-phantom barotropic fluids
excluded at the $95.4\%$ confidence level based on the earlier data
set. We notice that there was a time the observations favored
$w(z=0)\leq-1$~\cite{Riess:2006fw} but now the observations favor
$w(z=0) \geq -1$. However, the conclusions are drawn at the $68.3\%$
confidence level at most. It is hoped that the next-generation
observations will constrain the dark energy equation of state an
order of magnitude better~\cite{Frieman:2008sn,Albrecht:2006um}. We
shall be able to identify dark energy at higher confidence in the
coming future.

\section{Acknowledgment}
C.-W.~Chen is supported by the Taiwan National Science Council under
Project No. NSC 95-2119-M-002-034 and NSC 96-2112-M-002-023-MY3, P.
Chen by the Taiwan National Science Council under Project No. NSC
97-2112-M-002-026-MY3 and by US Department of Energy under Contract
No. DE-AC03-76SF00515, and Gu by the Taiwan National Science Council
under Project No. NSC 98-2112-M-002-007-MY3. All the authors thank
Leung Center for Cosmology and Particle Astrophysics for the
support.

\end{document}